\pdfoutput=1
\documentclass{JINST}

\usepackage{textcomp}
\usepackage{float}
\usepackage{ifpdf}

\title{Versatile firmware for the Common Readout Unit (CRU) of the ALICE experiment at the LHC}

\author{O. Bourrion\thanks{Corresponding author.}, J. Bouvier\\
Univ. Grenoble Alpes, CNRS, Grenoble INP\textsuperscript{$\dagger$}, LPSC-IN2P3, 38000 Grenoble, France \\
$\dagger$ Institute of Engineering Univ. Grenoble Alpes
}

\author{F. Costa\\
        CERN 1211 Geneva 23 Switzerland}

\author{E. D\'avid, J. Imrek, T.M. Nguyen\\
          Wigner Research Centre for Physics, 29-33 Konkoly-Thege Mikl\'os str,\\
          H-1121 Budapest, Hungary}
        
\author{S. Mukherjee\\
        Bose Institute, Kolkata 700 054 WB, India}       
        
\abstract{
As from the run 3 of CERN LHC scheduled in 2022, the upgraded ALICE experiment will use a Common Readout Unit (CRU) at the heart of the data acquisition system.
The CRU, based on the PCIe40 hardware designed for LHCb, is a common interface between 3 main sub-systems: the front-end, the computing system, and the trigger and timing system.
The 475 CRUs will interface 10 different sub-detectors and reduce the total data throughput from 3.5 TB/s to 635 GB/s. 
The ALICE common firmware framework
supports data taking in continuous and triggered mode and forwards clock, trigger and slow control to the front-end electronics.
In this paper, the architecture and the data-flow performance are presented.
}

\begin{document}


%
%
\section{Introduction}
The ALICE upgrade addresses the challenge of reading out Lead--Lead (Pb--Pb) collisions at rate of 50\,kHz, proton--proton (pp) and proton--Lead (p--Pb) at 200\,kHz and higher. 
This will result in the collection and inspection of a data volume of heavy-ion events roughly 100 times larger than during Run 1 and 2.
From Run 3 on, the majority of ALICE sub-detectors are upgraded to operate in continuous, trigger-less readout mode. 
This a consequence to the fact that a very low signal-to-background ratio is expected in the low--$p_T$ region, as the rate of collisions of interest will be of the same order as the interaction rate \cite{aliceTDR}.
However, triggered readout is still used by all detectors for commissioning and calibration runs. Also, some detectors which are not upgraded still use the legacy readout systems in triggered mode. 
The 13 ALICE sub-detectors read out via $\approx$ 10'000 readout links produce 3.5\,TB/s of data. In order to cope with the continuous readout and the resulting data throughput 10 detectors have been upgraded to use the Common Readout Unit (CRU).
Faithful to the design reuse strategy for the LHC experiments, the PCIe40 electronics designed for the LHCb experiment \cite{Pcie40publi} is used as the ALICE CRU.
The CRU is a PCIe-gen3 based FPGA processor board with up to 48 bidirectional optical links (48 in, 48 out).

\begin{figure}[hbtp]
\centering
\includegraphics[angle=0,width=0.8\textwidth]{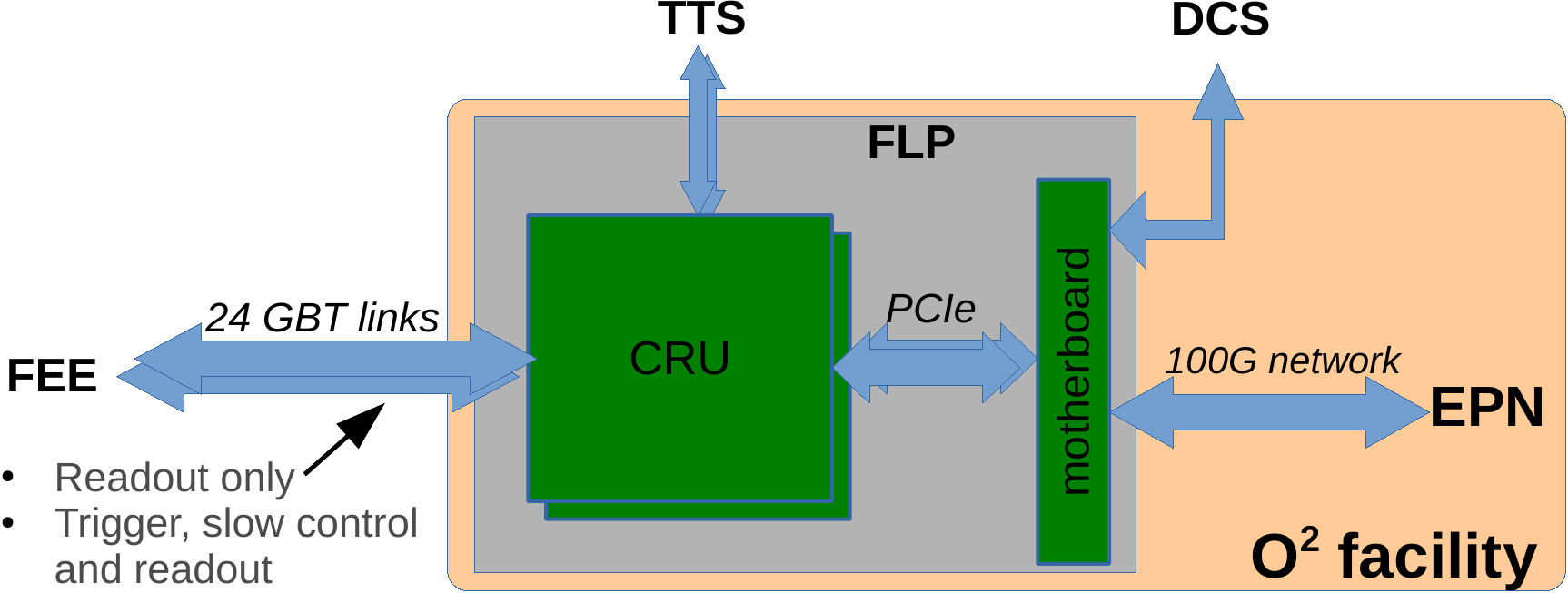}
\caption{\label{cruInSyst} The CRU is the interface between the detector front-end electronics, the O\textsuperscript{2} facility, the DCS and the TTS via the Local Trigger Unit (LTU).}
\end{figure}

The CRU is the interface between the front-end electronics (FEE), the Online-Offline facility (O\textsuperscript{2}), the Detector Control System (DCS) and the Trigger and Timing System (TTS), see Fig.\,\ref{cruInSyst}. 
One CRU can interface up to 24 optical links with the front-ends, one trigger link with the TTS and one PCIe interface.
The GBT protocol \cite{GBTpubli,GBThtml} developed by the electronics is used to communicate with the front-end electronics.
In ALICE it was chosen to aggregate a limited number of links (24) to limit the risk to loose a large portion of some detectors in case of hardware failure.

The O\textsuperscript{2} facility is a computer farm composed of the First Level Processors (FLP) and Event Processing Nodes (EPN). 
The FLP (DELL POWEREDGE R740) exchanges information with the FEE via the CRU, it can host a maximum of three CRUs. 
The FLP communicates with the EPN through a 100\,Gb InfiniBand network. 

The front-end interface, ensured via the GBT link, can be used to receive/deliver the following information:

\begin{itemize}
  \item READOUT (PHYSICS data), from FEE to CRU.
  \item TRIGGER and TIMING (clock and trigger information), from CRU to FEE.
  \item SLOW CONTROL, both directions.
\end{itemize}

The communication with the TTS is done via a dedicated bidirectional Passive Optical Network (PON) \cite{ttcpon}.
The PON  is time multiplexed in the upstream direction (CRU toward TTS).
It allows the reception of the LHC machine clock and of trigger and decision messages in the downstream direction (TTS toward CRU) and acknowledge messages in the upstream direction.

The CRU is connected to the server's motherboard via dual PCIe gen3 x8. The DMA is dedicated to the readout of the detector while control messages from the DCS are passed through the Base Address Register (BAR) interface.

\section{Hardware overview}
\begin{figure}[hbtp]
\centering
\includegraphics[angle=0,width=0.95\textwidth]{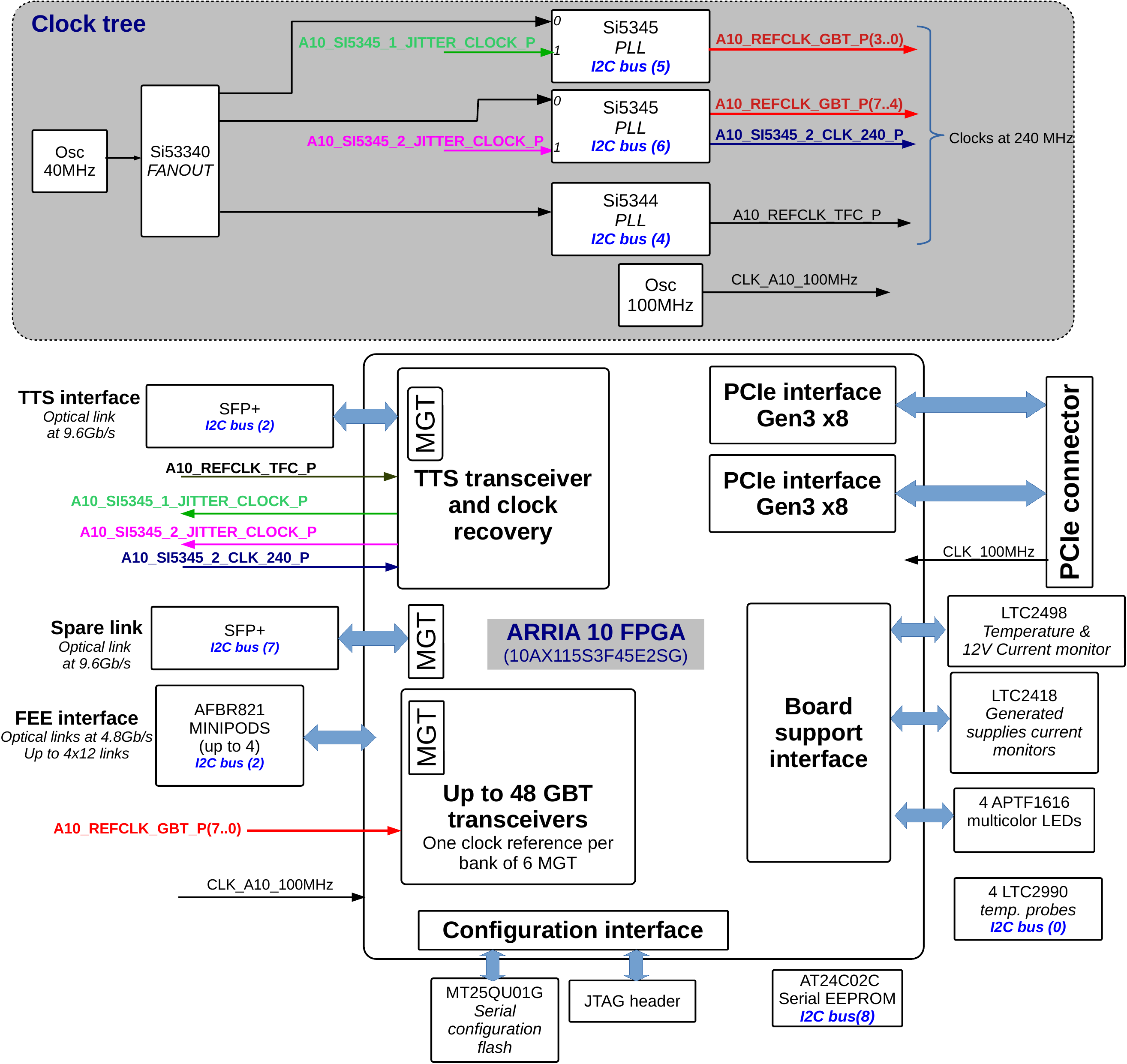}
\caption{\label{hardwareFig} CRU hardware overview. The clock tree is shown as well as the FPGA and its interfaces with the various components of interest. The TTS and the GBT transceivers use dedicated Multi Giga Bit transceivers (MGT).
}
\end{figure}

A functional overview of the hardware highlighting the features used in ALICE CRU can be seen in Fig.~\ref{hardwareFig}. The clock tree is shown as well as the FPGA and its interfaces with the various components of interest.

The clock tree is designed such as to use a common reference for all communication links of the CRU, aside from the PCIe which uses 100\,MHz.
The board can be operated either in standalone mode with a 40\,MHz oscillator available on the board or from a recovered clock extracted from the optical connection with the TTS. 

However, the TTS transceiver needs a stable 240\,MHz reference clock at startup, locally generated by a SI5344PLL.
Once locked to the incoming stream, the recovered clock is sent from the FPGA to a high performance PLL (SI5345) for jitter attenuation.
The cleaned clocks are then used to operate the FPGA logic and the GBT.
The selection between the local and recovered clock mode is done with the SI5345 PLL via I2C communication.
A free running 100\,MHz clock is produced on board and used to operate the miscellaneous functions embedded within the FPGA, such as initialisation and hardware monitoring.

The FPGA is an ARRIA10 from Intel (10AX115S3F45E2G). 
It is connected to two Small Form Pluggable (SFP+) connectors. 
Only one SFP of the two is used for the TTS connection, the second is a spare connection. 
Up to 4x12-channel bi-directional 10.3125\,Gb/s  optical transceivers (mini-pods \cite{minipods})  ensure the connections to the front-end electronics via GBT links. 

For ALICE, only two mini-pods are equipped to allow the connection to 24 front-end links with the only exception being the TRD detector, which does not use the GBT protocol and where 36 links and 3 mini-pods are needed.

On the back-end side the CRU is connected to the PCIe edge connector and offers a dual gen3 x8 PCIe interface.
This interface is clocked by the reference clock of 250\,MHz provided through the connector.

The FPGA is also connected to board support functionalities such as temperature and current sensors, and an EEPROM which contains a unique identifier set by the manufacturer during board assembly.
Communication to these peripherals is via I2C or Serial Peripheral Interface (SPI).
There is also the possibility to control multi-color LEDs which is useful when trying to quickly locate a specific machine in a server farm for maintenance purpose.

Finally, the FPGA can be configured either from a JTAG probe, which is handy for debugging the firmware in lab, or from a Quad SPI flash.
The latter can be reconfigured remotely via the PCIe interface, allowing on-site upgrades.
%
%
\section{Data readout}
\begin{figure}[hbtp]
\centering
\includegraphics[angle=0,width=0.8\textwidth]{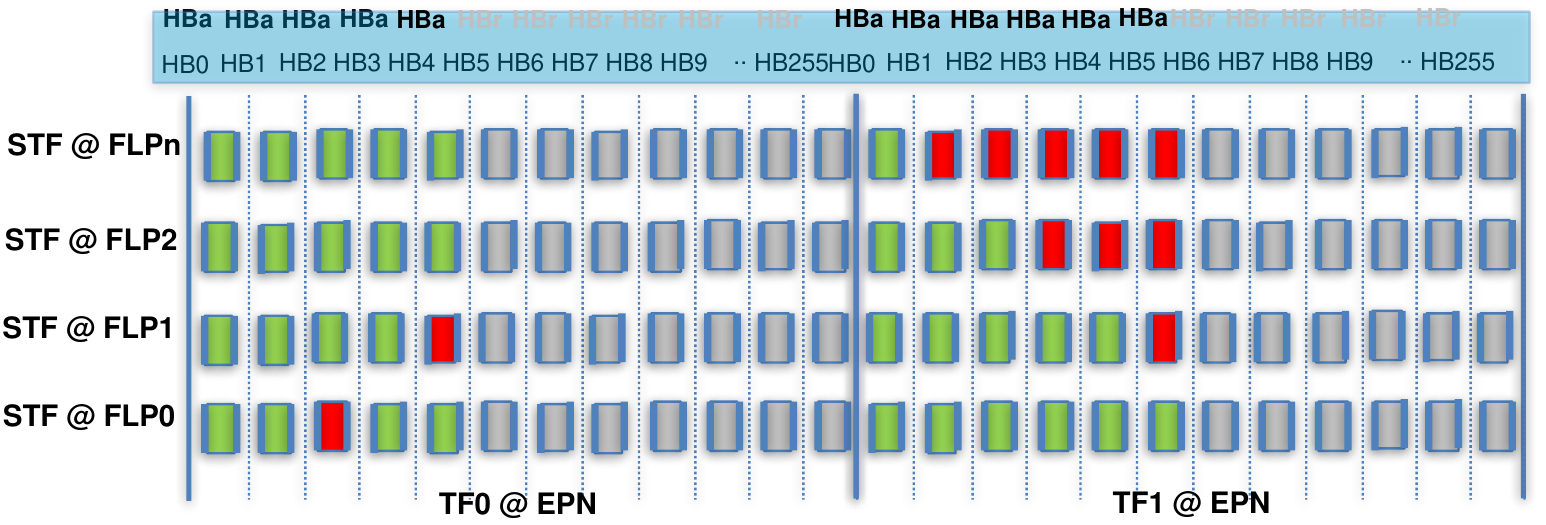}
\caption{\label{continuousFig} Illustration of the continuous readout (extracted from \cite{run34note}). 
Each rectangle represents a Heart Beat Frame (green: successfully received; red: bad reception or missing fragment; grey: deleted HBF).
There are up to 256 HBFs per Sub Time Frame (STF) produced by each FLP and the Time Frame is the collection of STF at the EPN level.
In this example, the remaining HBFs of a TF are rejected after a bad reception is detected.
HBa: Heart Beat accept; HBr Heart Beat reject.
}
\end{figure}

The ALICE computing upgrade concept consists of transferring all detector data unfiltered (triggerless) to the computing system. 
Data volume compression will be performed by processing the data on the fly and not by rejecting complete events as do the high-level triggers or event filter farms of most high-energy physics experiments.
For event reconstruction at the EPN level, the continuous data stream is sliced in Time Frames (TF) of programmable length of maximum of 22\,ms. 
The Times Frames are then divided into Heart Beat Frames (HBF) of one orbit duration (89.4\,\textmu s).
Heart Beat and Time Frame triggers indicating the boundaries between HBFs and TFs are distributed to the CRU via the PON network by the Central Trigger Processor (CTP).
In this scheme, the task of the CRU is to collect the data continuously and to check the successful Heart Beat Frame transmission to each First Level Processor. 
The CTP distributes Heart Beat triggers defining whether the corresponding HBF should be accepted (HBaccept - HBa) and thus forwarded to the FLP or deleted (HBreject - HBr).
This scheme allows the data throughput to be adjusted to match the available bandwidth during commissioning and dedicated calibration runs.
For each HBF, each CRU delivers \textit{an acknowledge} (HBACK) or \textit{not acknowledge} message (HBNACK) to CTP which assesses the quality of the HBF transmission of all CRUs.
In the case incomplete or corrupted HBFs have been transmitted, the CTP can request the corresponding HBF or even those from a full TF to be deleted from the FLP memory.
An example of two incomplete time frame transmissions is shown in Fig.\,\ref{continuousFig}.

The upgraded detectors are classified either as streaming or as packet detector types and can be read out via the CRU in two readout modes: continuous and triggered.
Both readout modes are supported by the two types of detectors, and the experiment will run in one of the two readout modes with a combination of different detector types.

\begin{itemize}
  \item Packet detectors can generate data formatted in packets before shipping them to the CRU. Every packet, so called GBT packets, contains the time stamp defined as orbit and bunch crossing.
  \item Streaming detectors cannot organize the data in formatted packets, therefore they generate a stream of data, so called GBT stream, without orbit and bunch crossing information.
\end{itemize}

As described before, data are transferred to the FLP memory in Heart Beat Frames (HBF). 
The data coming from the different detectors reach the FLP memory with an identical format.

\subsection{GBT packet type}
The CRU expects to receive HBFs prepared in the FEE and transferred over the GBT links. 
The HBF consists of one or more blocks of data containing one header, called Raw Data Header (RDH) and the payload. 
Data coming from detectors delivering GBT packets are forwarded to the DMA engine unmodified by the CRU firmware. 
The CRU verifies only the correct HBF structure and formatting. 
For every HBF the CRU sends an acknowledge message to the CTP informing whether the corresponding HBF has been correctly received (data received properly from all links involved in the data taking) and forwarded to the FLP.

\subsection{GBT stream type}
GBT stream type sequences do not have an HBF structure and consist of a continuous stream of detector specific raw data entries. 
Upon receiving HB triggers, the CRU partitions the data stream in HBFs and forwards them to the DMA engine or deletes the corresponding data packets. 
In this configuration each GBT link generates a data throughput of up to 4.48\,Gbps, regardless the content of the data.
In normal operation the detector specific user logic of the CRU performs zero suppression in order to reduce the data throughput.

For short debugging and calibration runs the data compression can be switched off to transfer sequences of uncompressed data, otherwise the amount of data would be too high to be handled by the rest of the data taking chain.
For the streaming detectors to collect data in the Physics runs, it is mandatory to implement the User Logic in the CRU, as described below.

%
%
\section{CRU firmware requirements}

The CRU firmware is divided into two parts. 
The first part is the common firmware which (i) provides the interfaces to PCIe, trigger and timing, and up to 24 front-end links via the GBT protocol, (ii) provides the possibility to read out all detectors in `raw-mode' with no data processing in the CRU, (iii) allows reference clock and trigger signals distribution, and (iv) permits FEE configuration.
The second part is the user logic which is only needed for those detector systems that need detector-specific data processing for instance baseline correction or zero suppression. 

An important ALICE requirement is to be able to switch between raw-mode and user-logic at any moment without reloading different firmware versions.
Self-testing capabilities to ease commissioning and system maintenance are implemented.
They are detailed later in this paper.

From a system point of view, the different requirements for the common firmware are due to the different GBT bus mode (packet or stream), the information it is supposed to carry, the integration of a user logic or not, and the type of slow control protocol required to configure the FEE.

 It is important to mention that when the data reach the DMA engine in the CRU the format is identical, regardless of the data taking configuration of the card.
 Having the same data format simplifies the firmware logic of the CRU as well as the software required for physics analysis.

The GBT links are  used to send downstream trigger messages and/or the reference clock to the FEE, while upstream they are used for data readout and optionally to acknowledge specific slow control transactions.
The CRU firmware supports operation of the GBT links in GBT-mode (80 bits of payload and 32 bits of forward error correction) or in wide-bus (112 bits of payload and no forward error correction).
For almost all the GBT detectors the GBT-mode provides sufficient data bandwidth. 
Only the Time Projection Chamber (TPC), pushes data using the 'wide-bus' mode in order to satisfy the bandwidth requirements.
As the radiation load for the TPC front-end cards is sufficiently low, the forward error correction the GBT-mode would provide is not required. 
The requirements of the various detectors can be summarized in table shown in table\,\ref{reqTable1}.

\begin{table}[hbtp] 
\centering
\includegraphics[angle=0,width=0.85\textwidth]{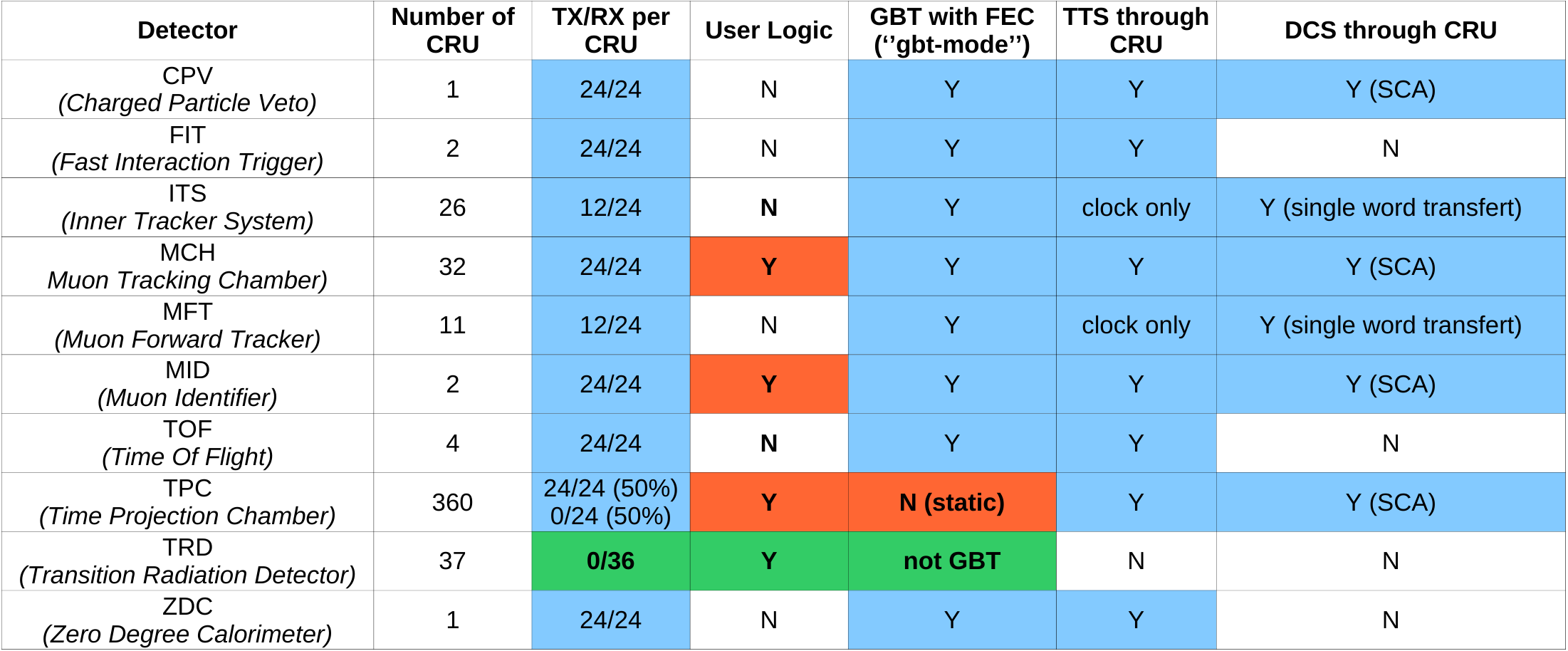}
\caption{Table summarizing the requirements of the various detectors. In white features not required; in blue compatible requirements with an adapted common firmware; in green and orange features requiring a specific firmware file generation.
}
\label{reqTable1}
\end{table}

%
%
\section{Firmware description}

\begin{figure}[hbtp]
\centering
\includegraphics[angle=0,width=0.75\textwidth]{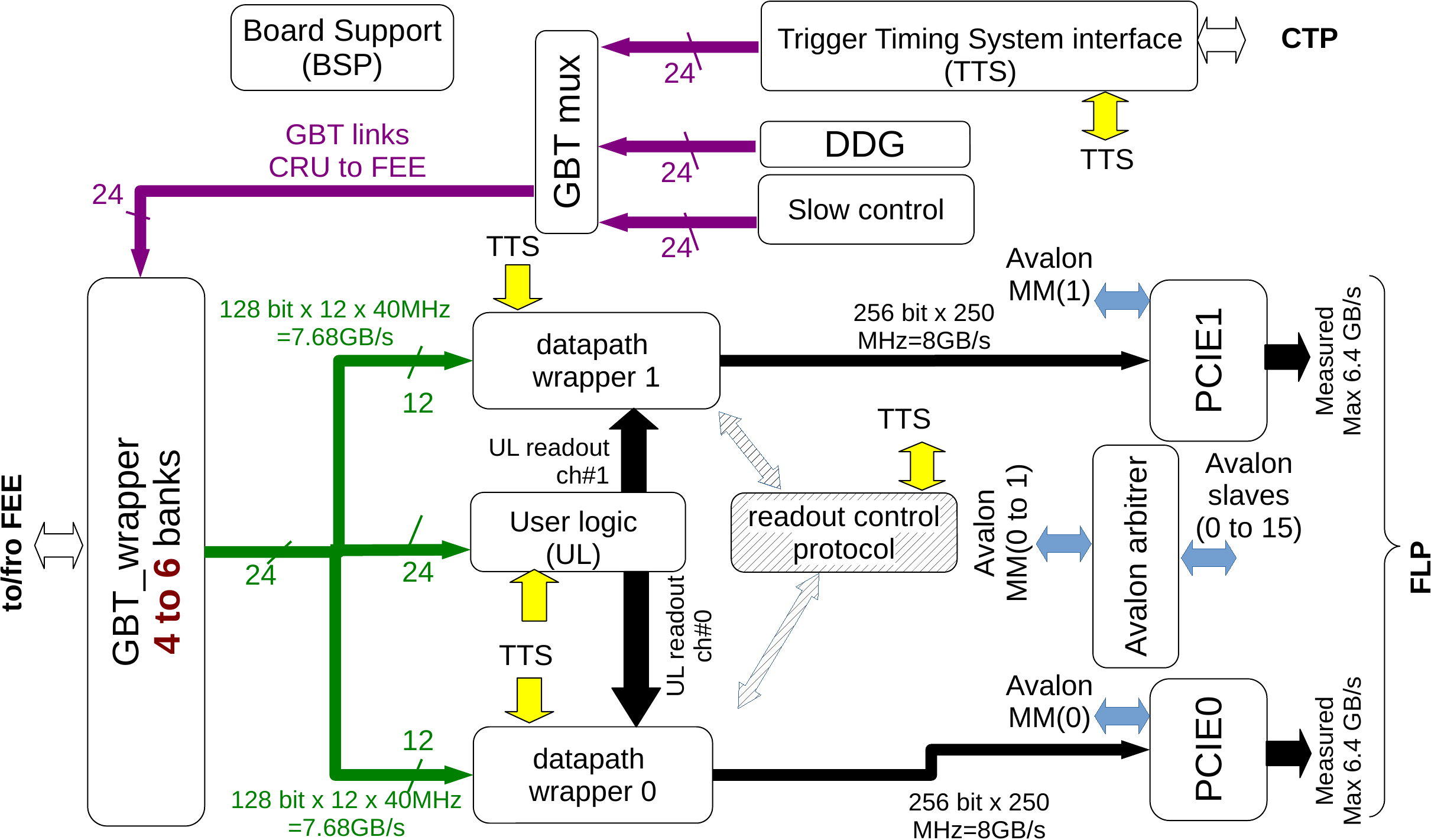}
\caption{\label{firmOverview} Overview of the common firmware. The main parts are shown: the GBT wrapper, the Trigger and Timing System interface, the data path and the PCIe endpoints.
}
\end{figure}

An overview of the firmware is given in Fig.\,\ref{firmOverview}. 
The main parts are shown: the FEE interface (through GBT links), the Trigger and Timing System (TTS) interface, the board support Package (BSP), the data path and the PCIe endpoints.
Starting from the front-end side on the left, the \texttt{GBT\_wrapper} interface is shown; it is the interface with the FEE.

On the downstream path (CRU to FEE), depending on the detector requirements or test requirements, several sources can be selected to supply the \texttt{GBT\_wrapper}.
These are the Trigger and Timing System interface, the Dedicated Data Generator (DDG) or the slow control.

\subsection{Board Support Package (BSP)}
The board support package features several I2C masters that are directly controlled through the PCIe interface. 
They allow the readout of the various optical transceiver parameters such as temperature or optical power, the settings of external PLLs and the access to the board serial number stored in an EEPROM.
Additionally, it permits accessing to the FPGA serial number (fixed by the FPGA manufacturer) and monitoring of the FPGA die temperature.

The BSP functionality also includes the reconfiguration of the QSPI flash and the possibility to trigger an FPGA reboot from the slow control (through PCie interface).
The chosen strategy is to have two reserved areas in the flash memory, one for a golden and one for the application firmware.
The golden firmware will never be modified outside of the lab, while the application firmware can be modified when deployed on site.
At cold startup, the FPGA always boots on the golden firmware.
Then, by issuing a PCIe BAR command, specifying the memory offset to use, it is possible to load the application firmware in the FPGA.
In the case of a configuration failure (e.g. loss of connection, power cut or faulty firmware) the CRU can be easily recovered.

\subsection{GBT wrapper}
\begin{figure}[hbtp]
\centering
\includegraphics[angle=0,width=0.75\textwidth]{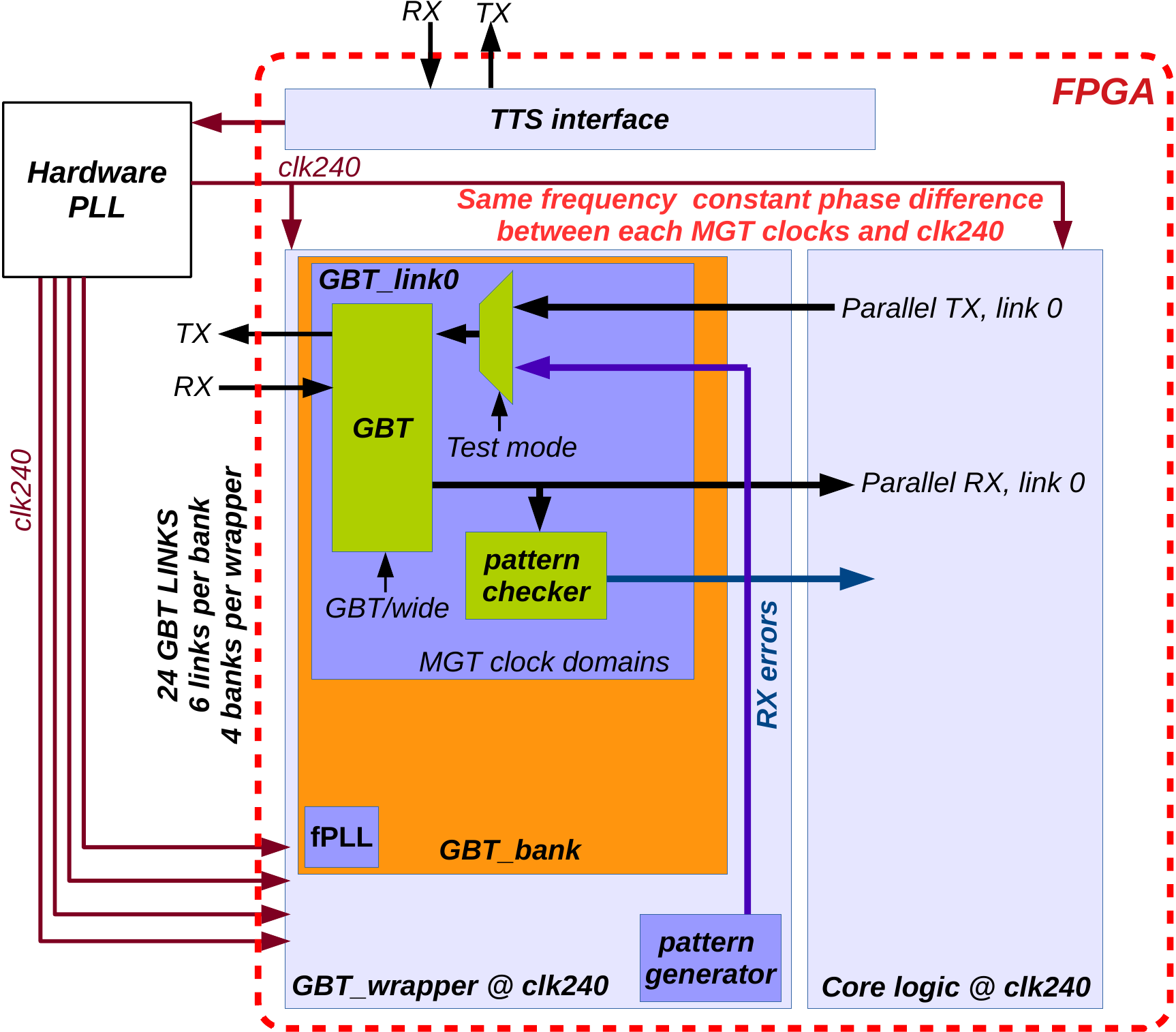}
\caption{\label{GBTfig} Modified GBT-FPGA inserted in the common firmware overview.
It shows the clock recovered from the Trigger and Timing System transferred to an external jitter cleaner PLL. 
The cleaned clock is re-injected into each bank of the FPGA used for the GBT connection. 
A single GBT link within a MGT bank is shown.
The pattern generator is shared between all links and operates with the common core clock (clk240)."
}
\end{figure}
The \texttt{GBT\_wrapper} is a modified version of the GBT-FPGA developed at CERN \cite{GBTpubli,GBThtml}, see Fig.\,\ref{GBTfig}.
The main differences are that (i) it has a user data path operating at 240\,MHz (six times the machine clock), (ii) the clock domain crossing between the transceiver domain and the user domain is achieved with timing constraints instead of a phase scan at link start-up, (iii) dynamic switching is possible between GBT-mode and wide-bus mode to cover all detector requirements and (iv) the test data pattern generator is shared between all links to save resources.
Moreover, the \texttt{GBT\_wrapper} permits external (with optical fibers) and internal (inside the FPGA transceivers) loop-back tests which allow the validation of the CRU-FEE communication and the CRU data path operation once installed in the system.
In external loop-back the data generator enables the emission of representative data towards the FEE that can be looped back by them into the CRU; while in internal loop-back mode, this feature allows to stress the system without relying on the availability of detector FEE.

The strategy used to maintain a constant latency and to avoid the phase scanning on the transmission path (CRU to FEE) is (i) to rely on the zero delay buffer provided by the hardware PLL and to feed the extracted transmission delays due to the PCB in the constraint file, and (ii) to use the 6 time faster rate to properly sample and transfer the data from one clock domain to the other (120 bits transferred at 40\,MHz).
On the receiving, which is used only for readout, a non constant latency can be accepted, and thus a FIFO was implemented to cope with the clock domain transfer from the recovered clock domain to the user part clock domain.
This solution was extensively tested across several scenarios and proved to be reliable. The scenarios tested were: CRU reference clock switching between local and remote clock source, warm reboot (FPGA reconfiguration) and cold reboot (CRU and FLP turned off and rebooted).

\subsection{TTS interface}
\begin{figure}[hbtp]
\centering
\includegraphics[angle=0,width=0.6\textwidth]{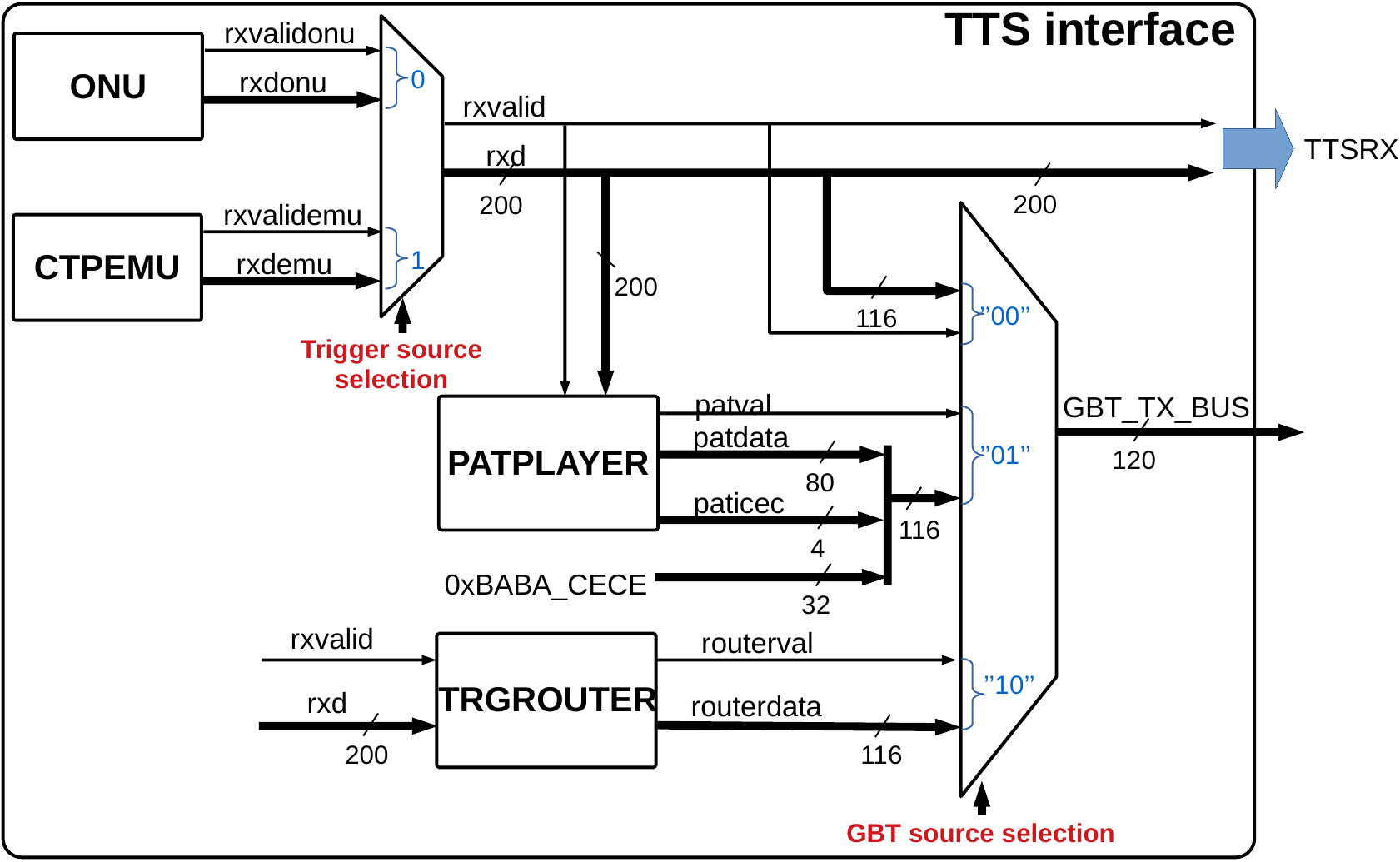}
\caption{\label{TTCfig} Overview of the TTS interface. The 200 bit word contains the trigger information on its lower 116 bits and the trigger decision message on its 80 bits upper part.
}
\end{figure}

As shown in Fig.\,\ref{TTCfig}, the TTS interface is composed of four components. 

The first is the Optical Network Unit (ONU) \cite{ONU1,ONU2} which recovers the machine clock from the PON and forwards it through the GBT to the FEE.
The ONU is also used to receive the trigger and timing message at each clock cycle (trigger bits, bunch crossing number, Heart Beat ID) from the central system.
The 200 bit word contains the trigger information on its lower 116 bits and the trigger decision message on its 80 bits upper part.
The CRU uses the upstream direction to send the HBACK or HBNACK trigger message.
As the optical network is passive, the upstream communication is time multiplexed and a message can only be sent every 125\,ns times the total number of ONUs connected on the PON.
The second component is a trigger emulator (ctpemu) that is used for tests and system diagnostic purposes. 
It can produce trigger messages, like the ones provided through the ONU, and simulate readout flow control by producing HBa and HBr commands.
The third component is the pattern player (patplayer) that can generate a programmable sequence to be transmitted to the FEE, it is started by a trigger bit issued either by the ONU or ctpemu.
The fourth is the trigger router (trgrouter) which remaps, duplicates and forwards some trigger bits received via the PON from the CTP to the FEE boards via the GBT links.

\subsection{Detector Data Generator (DDG)}
The DDG is the component in the CRU able to emulate detector behavior and data throughput.
This component plays a main role during the test and validation of the CRU firmware and the readout software chain when the detector hardware is not yet widely available. 
The DDG has different configuration parameters and it can be dynamically configured to produce either streaming or packet type data. 
The data packet can be produced for different trigger types with fixed or random packet length and inter packet duration, generating a realistic detector throughput.
The DDG can be used to test the firmware at any moment in time. 
Configuring the GBT links using the internal loop-back connection makes the injection of DDG data in the system possible without the requirement to change the physical optical connection at the card input.
DDG is a powerful self-test feature to verify the correct behavior of the hardware and the software without relying on external hardware elements like the FEE. 

\subsection{Slow Control}
The majority of the detectors are connected to the DCS system through the CRU. 
In ALICE there are three configuration protocols:

\begin{itemize}
  \item GBT-EC (External Control)\cite{GBThtml}, used to send configuration data to the GBT SCA ASIC (Slow Control Adapter)\cite{scaRef} installed on the FEE.
  \item GBT-IC (Internal Control)\cite{GBThtml}, used to configure the GBTx ASIC on the FEE.
  \item GBT-SWT (Single Word Transfer), not using the GBT protocol.
\end{itemize}

While the GBT-EC and IC protocols are part of the GBT design developed at CERN, the SWT is ALICE specific and can be used only by detectors that host an FPGA in the FEE, see Table~\ref{reqTable1}.
The SWT protocol has been introduced to increase the bandwidth for the slow control operation.
It uses the GBT-DATA path to deliver up to 3200\,Mb/s (80\,bits per 40\,MHz clock cycle) to the FEE whereas the GBT-EC path provides 80\,Mb/s.
In practice the slow control read/write speed is limited to 36\,Mb/s by the time taken for software to access the PCIe BAR.
To send the detector configuration data over the GBT link using the SWT protocol, the CRU must be configured to switch to the SWT traffic with the GBT-MUX component (see Fig.~\ref{firmOverview}). 
This is a static selection, as there is no dynamic packet switching for the downstream path.
In the opposite direction, from the FEE towards CRU, the SWT and FEE information is interleaved within the same link.

In order to distinguish GBT words that contain physics information from the control words, like the SWT, the CRU uses two types of information, the \textit{isdatasel} flag decoded from the GBT header and a part of the GBT word embedded in the GBT data field.

\begin{figure}[hbtp]
\centering
\includegraphics[angle=0,width=0.6\textwidth]{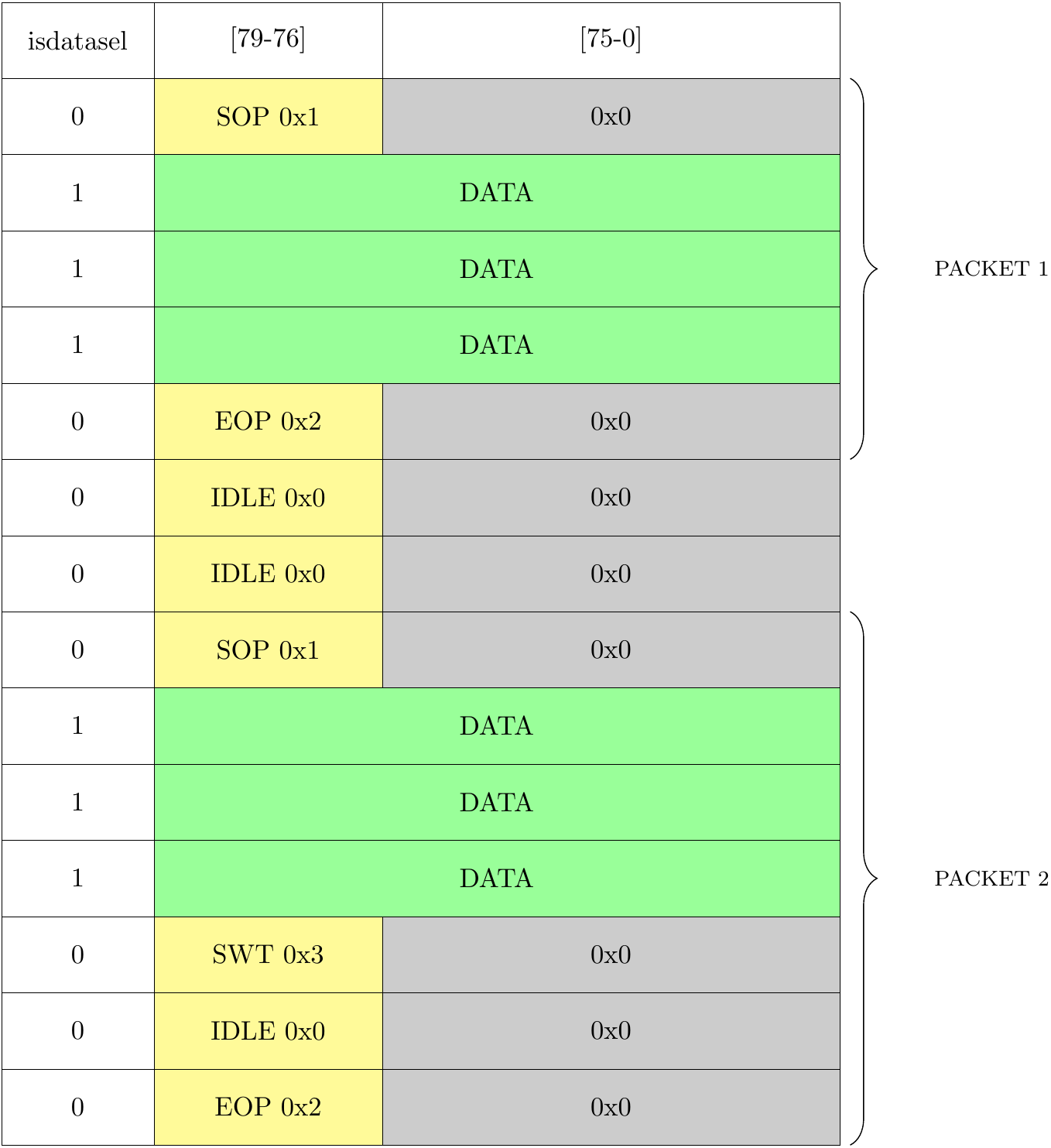}
\caption{\label{gbtdatafig} GBT stream from a packet detector FEE to a CRU. The physics word are marked by the flag \textit{isdatasel} set to 1 and the control words: IDLE, SOP, EOP, SWT are marked by the flag set to 0 and with a different header stored in the GBT data field.
}
\end{figure}

When the detector sends physics data the flag \textit{isdatasel} is set to 1 and the whole 80-bit data field is used to transfer the information. 
When this flag is 0 the CRU considers the GBT word a control word and it uses the four most significant bits of the data field to distinguish between the different control words (Fig.~\ref{gbtdatafig}):

\begin{itemize}
    \item IDLE: 0x0. This word is used to pause the data transfer in case the FEE has no data to send to the CRU.
    \item SOP: 0x1. This word identifies the Start Of Packet during a GBT data transfer from a GBT packet type detector.
    \item EOP: 0x2. This word identifies the End Of Packet during a GBT data transfer.
    \item SWT: 0x3. Single Word Transfer, it is used to identify a configuration word.
\end{itemize}

The CRU extracts the SWT information from the data stream before it reaches the DMA engine and stores it in a dedicated FIFO which is accessed by DCS.

\subsection{Datapath wrapper}
\begin{figure}[hbtp]
\centering
\includegraphics[angle=0,width=0.75\textwidth]{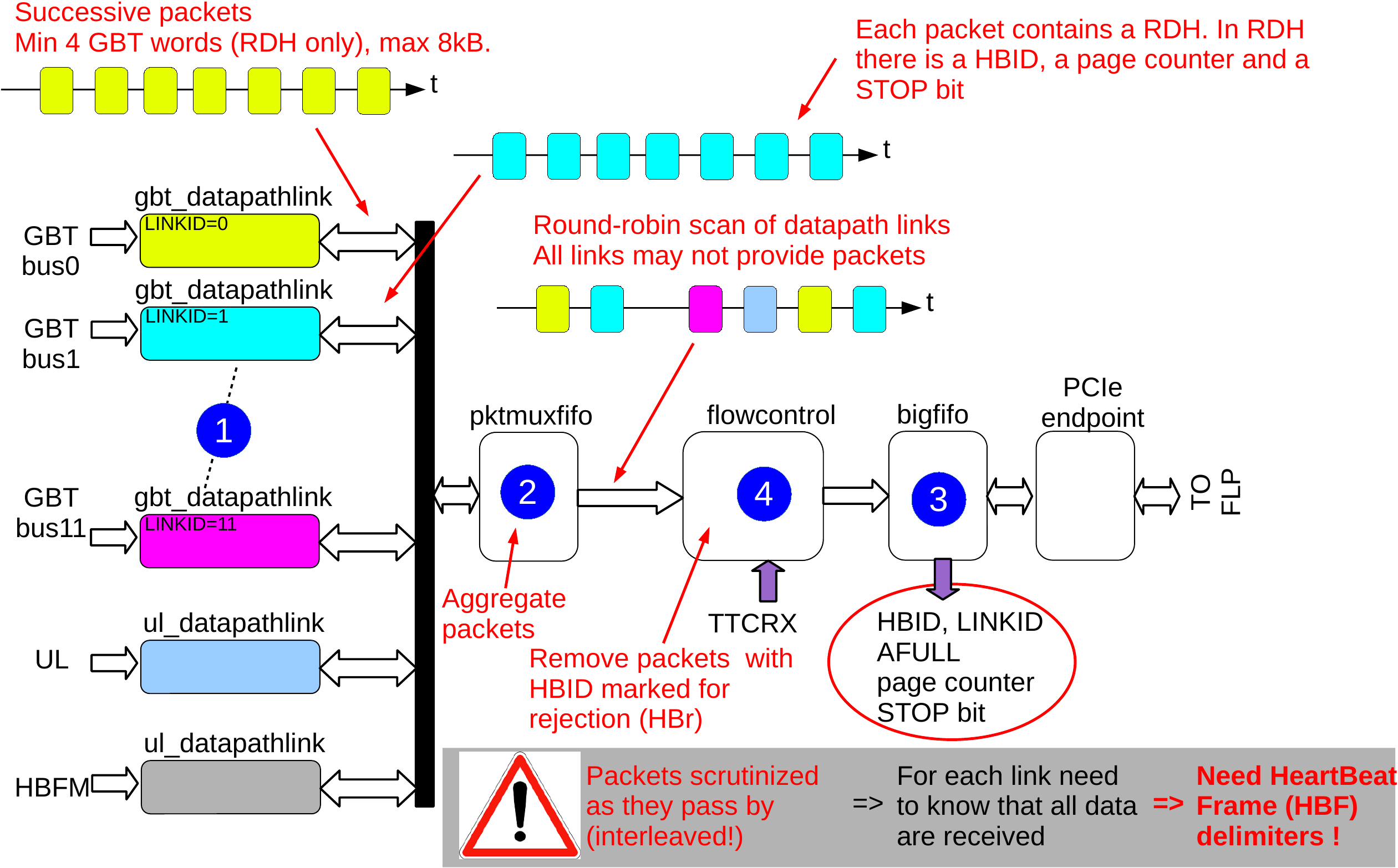}
\caption{\label{dwrapperOverview} Overview of the \texttt{datapath\_wrapper}.}
\end{figure}

Two \texttt{datapath\_wrapper} blocks are implemented in the firmware.
They receive the trigger messages, collect and aggregate the FEE data provided by the GBT link or use directly the \texttt{user logic} input if implemented. 
They also provide monitoring information to the \texttt{readout control protocol} component (see  Fig.\,\ref{dwrapperOverview}).

The first task of each \texttt{datapath\_wrapper} is to receive in parallel the data either from up to 12 GBT busses, from the \texttt{readout protocol} component (trigger acknowledge or decision message) and/or from one user logic link.
The \texttt{gbt\_datapathlink} is compatible with stream or packet type format.
When selecting the stream mode this component constructs data packets, by chopping the data stream and inserting the Raw Data Header (RDH).
The RDH describes the readout packet content and contains the Heart Beat ID (HBID), the Link ID, the page counter and the stop bit as well as other information.
For each Link ID the page counter gives the packet identification within the corresponding HBF transmitted and the stop bit indicates whether the last packet for this HBF is being transmitted.
The \texttt{ul\_datapathlink} receives already correctly formatted packets, either from the User Logic or from the Readout Protocol block, i.e. Heart Beat Acknowledge and Decision Messages (HBFM).
At the output of this first stage the packets have a maximum size of 8\,KB.
The second stage, named pktmuxfifo, performs data aggregation by scanning in a round-robin based manner possible data sources (\texttt{gbt\_datapathlink} and \texttt{ul\_datapathlink}) and collects data packets. 
At the output of this stage, the packets from the various links are interleaved.
If required by the CTP, this is followed by the removal of all packets from the data flow with a HBr message (number 4 in Fig.\,\ref{dwrapperOverview}).
Then, the packets are stored in a large buffer (bigfifo, 16\,kwords of 256 bits)) to be made available to the PCIe endpoint.
While being stored, the packets are scrutinized and useful parameters (HBID, LINKID, FIFO status) are presented to the \texttt{readout control} protocol component.
The \texttt{readout protocol} uses the information provided by both \texttt{datapath\_wrapper} blocks to check the interleaved packets.
The HBF reception is declared successful only if for each LINKID included in the readout, start (page counter is 0 in RDH) and stop packets (stop bit is 1 in RDH) were received in consecutively and properly stored in the \texttt{bigfifo} buffers before a pre-defined timeout for reception elapsed.
Then a HBACK or HBNACK message is transmitted to the CTP, which assembles the messages from all CRUs and updates the HBa/HBr messages to communicate whether a given HBF should be maintained or be deleted in the FLP. \cite{run34note,run34paper}.

\subsection{PCIe DMA}
\begin{figure}[hbtp]
\centering
\includegraphics[angle=0,width=0.75\textwidth]{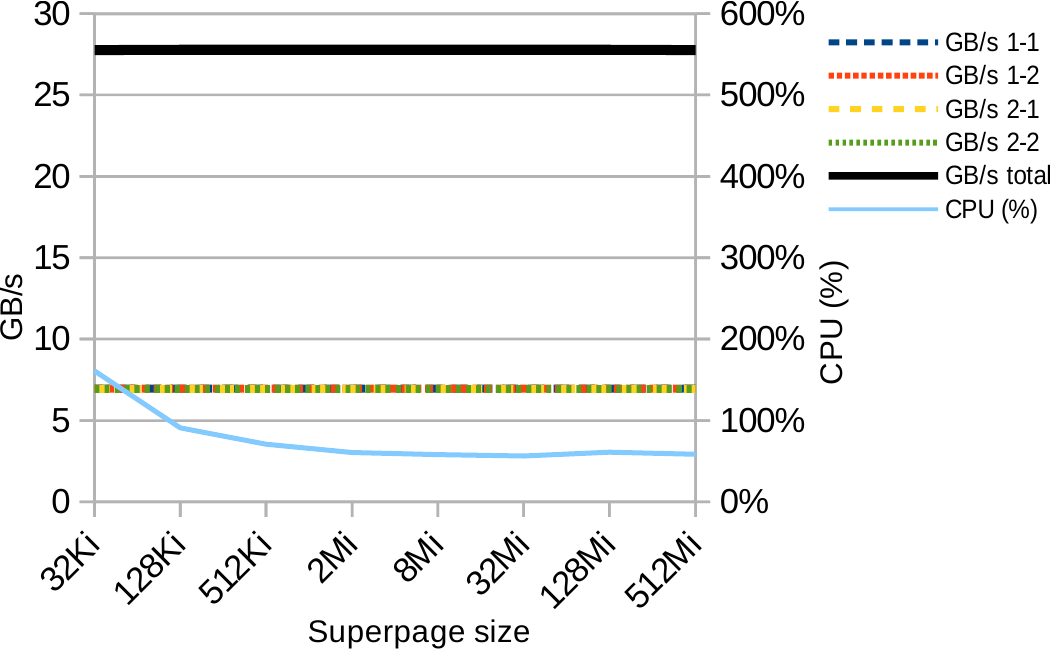}
\caption{\label{dmaperf} The graph shows the measured throughput performance of two CRUs installed in one server and of the CPU usage versus the super-page size. There are in total four endpoints, each one absorbing the maximum allowed data throughput of 55\,Gb/s. Note the CPU (Intel Xeon Silver 4210) usage decreases with increasing size of the super-page. }
\end{figure}

The main data stream flows through the DMA interface which is used to move data from the FEE to the FLP memory.
The CRU communicates with the FLP server through a PCIe gen. 3x16 interface implemented in the FPGA as a dual endpoint PCIe gen3 x8. 
The DMA engine of the CRU is capable of sustaining a total data throughput of 110\,Gb/s.
In order to achieve this performance, both endpoints must work in parallel, each one handling a maximum of 55\,Gb/s (Fig.~\ref{dmaperf}).

The nominal DMA throughput provides sufficient margin to collect data from the most demanding detector in ALICE, the TPC.
Each of the 360 TPC CRUs receives data from 20 GBT links as input for a total throughput of roughly 89.6\,Gb/s.
Unfortunately, the data must be aligned to 32 bit words boundary, and therefore the actual throughput is 102.4\,Gb/s.
The incoming data are compressed by the TPC user logic before being delivered to the DMA engine.
The expected output of the TPC user logic is 20\,Gb/s. 

In the firmware the CRU input data stream is divided between the two PCIe endpoints in order to avoid dynamic switching of the data flow between the two PCIe endpoints.
Thus, half of the GBT links implemented in the CRU are connected to one single endpoint via its datapath-wrapper instance that receives data from a maximum of 12 GBT links.
In this way, the data throughput is evenly distributed between the two PCIe gen3 x8 interfaces.

The communication with the software \cite{rocDriverRef} happens through the PCIe BAR interface.
There are two BAR interfaces: BAR0 and BAR2.
BAR0 is dedicated to DMA operations, it passes the page descriptors and monitors the status of the data taking, while BAR2 is used to access the card configuration and to monitor the other components of the firmware.
The CRU transfers the physics data into the server memory using DMA transfers.
In order to do so the CRU needs to know in which buffer data should be stored. 
For that, the software prepares (or allocates) the buffers in the FLP memory and then passes the addresses of each of these buffers to the CRU. 
This information is called the DMA page descriptors.
Therefore, to maximise the throughput by reducing wait states, the descriptors are prepared in advance by software and delivered through a FIFO to the CRU DMA engine.
If the FIFO holding the page descriptors becomes empty, the DMA transaction are paused and data are dropped in full DMA pages until new page descriptors are provided.

To reduce the software interaction with the hardware, the concept of super-pages has been introduced. 
A super-page is a buffer of contiguous spaces in memory of usually 1 or 2\,MB size (configurable based on the event size of the detector). The software stores the address of this buffer in the CRU descriptor FIFO and the DMA engine fills up the buffer with the data coming from the input links. Once the super-page is full a new descriptor is fetched and data is stored in a new super-page. It is the responsibility of the CRU firmware to divide the super-page in smaller DMA pages, of a maximum size of 8\,KB.

\begin{figure}[hbtp]
\centering
\includegraphics[angle=270,width=0.75\textwidth]{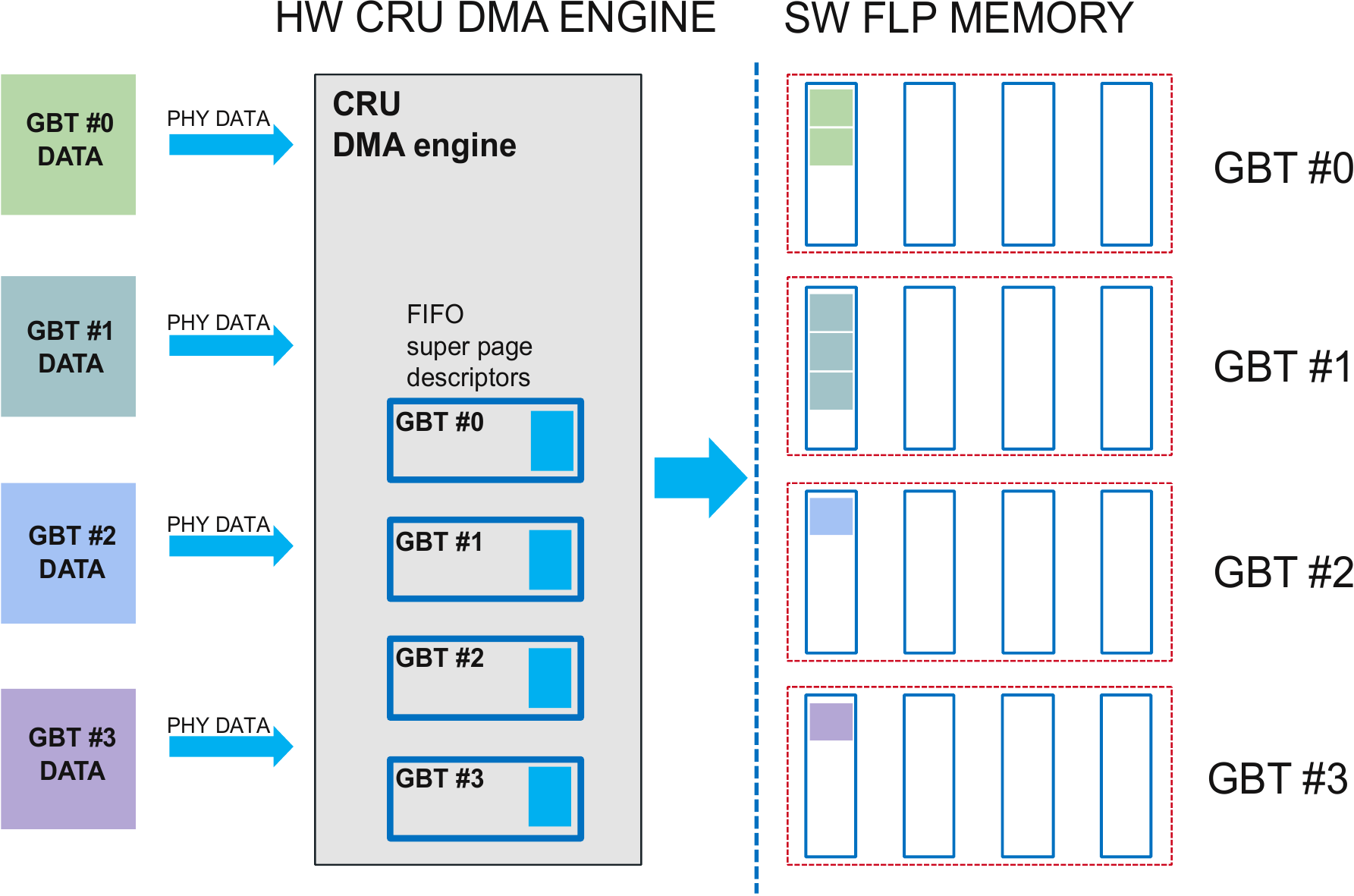}
\caption{\label{memflp} DMA with four GBT links. Data are moved into dedicated memory buffers.}
\end{figure}

The CRU can collect the data from a maximum of 24 GBT links. 
During data taking the CRU is loaded with 128 super-page descriptors for each GBT link active in the data taking.
Data coming from different links are organized in dedicated super-pages, therefore the software fetches data from specific GBT links in dedicated link memory buffers (as shown in Fig.~\ref{memflp}).

Data in the FLP memory is grouped before being sent to the EPN for further analysis. 
The grouping is done at the boundary of Time Frames. 
To simplify the job of the software the HBFs are stored by the CRU in dedicated super-pages. 
In this way it is simpler to group the different buffers belonging to a single Time Frame and send them over RDMA (Remote DMA) to the EPN farm. 
Assigning a dedicated buffer for each GBT link removes any dependency between them, improving the stability of the system and the total throughput.

\section{Firmware development, simulation and integration}
As the firmware implementation is quite complex and requires different configurations with different parameters depending on the sub-detector application a dedicated simulation strategy was developed.
The high-speed serial interfaces were simulated and validated on their own as they required much simulation power and would delay significantly a full simulation. 
Dedicated simulation test-benches were developed for the PCIe DMA interface and for the modified GBT-FPGA which was simulated against the reference design in its Xilinx implementation flavor.
The ONU interface was not simulated because it used an IP developed and validated by CERN.

The core of the design was embedded into a dedicated test-bench.
Simulation models emulated the data flow from the GBT-detectors (wide or GBT mode) while data sent to the PCIe DMA interface was recorded in a file.
The CTP messages were generated by the internal CTP emulator.
The FPGA Intel Avalon bus \cite{Avalon} required for configuration was connected to all the simulated components using it.
An Avalon master model was used to set the various registers during the simulation.
Thus, during the simulation, the configuration sequences required by the sub-detectors were simulated as they take place in real setups.
This allowed a reduction of the development time, but also the possibility to reproduce in simulation error cases that were detected by users.
Consequently, error resolution was faster, particularly for rarely occurring errors.
To ease the simulation, a hexadecimal address table was declared in a VHDL common package and used as reference for these configuration simulations.
Custom software extracted the address table information from the VHDL common package for use in the software running on the FLP.

The common firmware is distributed via a git repository to all users.
Makefiles included in the distribution allow users to easily simulate or compile the firmware.
A reference user logic generating known data is included in the reference design.
It can be configured to generate pre-defined and random packet sizes and data rates. 
This feature along with the DDG is a complementary tool for stressing in-situ the firmware and validating the readout software. 

In ALICE the CRU firmware deployment remains under the CRU team responsibility.
Sub-detectors add their user-logic code to a branch of the common firmware. The CRU team validates the code before integrating it into the master branch and finally deploying it.
The branch is then merged in the global repository, the firmware is compiled and delivered to the FLP in the ALICE system by the central team.

To keep track of the firmware after deployment and avoid misunderstanding, the firmware parts (common and user if any) are compiled after being committed in the repositories. In the provided scripts a routine generating tagging information is executed.
The generated VHDL parameters are included in the firmware and fed to status registers before actually starting the compilation. 
Thus after compilation, by accessing these registers, it is possible to retrieve the compilation date and the git hash of the compiled code.

\section{Resource usage}
A significant effort was made to find a good trade-off between the minimization of the FPGA resources and providing as much adaptability as possible to cover all detectors requirements, while providing embedded system test features.
The common part of the firmware uses about 123k (29\%) Adaptive Logic Module (ALM) and 1084 (40\%) RAM blocks of the available resources. 
The allocation of resources to the main functional blocks is shown in table~\ref{resourceTable} for the case where GBT links can be dynamically switched between GBT-mode and wide-mode.

\begin{table}[hbtp] 
\centering
\includegraphics[angle=0,width=0.95\textwidth]{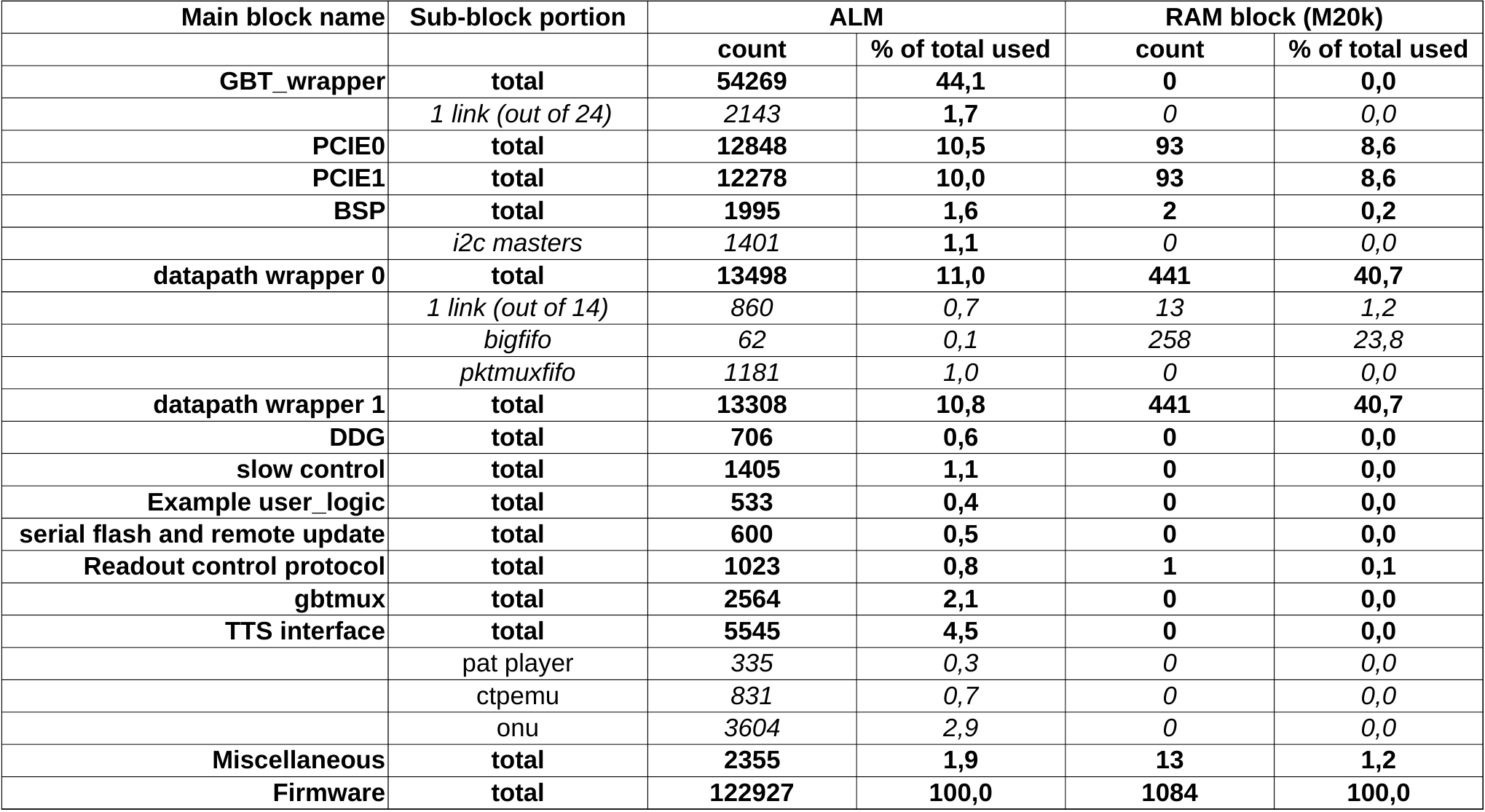}
\caption{Table summarizing the FPGA resource usage of the various firmware components and their relative contribution to the total.
}
\label{resourceTable}
\end{table}
From the table, it can be seen that the highest resource user is the GBT wrapper, which accounts for 44\% of the total. 
Fortunately, for the most demanding detectors in terms of resource usage for the user logic, which is TPC, the firmware can be compiled in wide-bus mode, thus removing the code error corrector at compilation time it is possible to save up to 30k ALMs.

\section{Conclusion}

An adaptable common firmware was developed to cover the needs of the upgraded detectors of ALICE \cite{gitProj}.
It was shown that by carefully designing the firmware features such as data path reader, CTP emulator, DDG, pattern player and by making them  configurable, it allowed the development and validation effort to be shared between the firmware and the associated readout software.
The different readout modes were validated by different detector setups in various running conditions.
At this moment 565 CRUs are produced. The board distribution is over and the installation is done. The validation of the detector specific firmwares, the ones featuring user logic, is in progress. The commissioning shall be over by the end of 2021.


\begin{thebibliography}{99}

\bibitem{aliceTDR} Upgrade of the Online - Offline computing system, ALICE collaboration \href{https://cds.cern.ch/record/2011297/files/ALICE-TDR-019.pdf}{CERN-LHCC-2015-006, ALICE-TDR-019}

\bibitem{Pcie40publi} J.P. Cachemiche \emph{et al}, \emph{The PCIe-based readout system for the LHCb experiment}, \mbox{\href{https://iopscience.iop.org/article/10.1088/1748-0221/11/02/P02013}{2016 JINST 11 P02013}}

\bibitem{GBTpubli} M. Barros Marin \emph{et al}, \emph{The GBT-FPGA core features and challenges}, \href{https://iopscience.iop.org/article/10.1088/1748-0221/10/03/C03021}{2015 JINST  10 C03021}

\bibitem{ttcpon} I, Papakonstantinou  \emph{et al}, \emph{Passive Optical Networks for the Distribution of Timed Signals in Particle Physics Experiments}, \href{http://dx.doi.org/10.5170/CERN-2009-006}{TWEPP09 proceedings, 	DOI:10.5170/CERN-2009-006}


\bibitem{minipods} Broadcom, \emph{AFBR-811 optical transceivers}, \href{https://www.broadcom.com/products/fiber-optic-modules-components/networking/embedded-optical-modules/minipod/afbr-811vxyz}{manufacturer website}

\bibitem{GBThtml} The GBT project website \href{https://espace.cern.ch/GBT-Project/GBT-FPGA/default.aspx}{https://espace.cern.ch/GBT-Project/GBT-FPGA/default.aspx}

\bibitem{scaRef} A. Caratelli \emph{et al}, \emph{The GBT-SCA, a radiation tolerant ASIC for detector control and monitoring applications in HEP experiments}, \href{https://iopscience.iop.org/article/10.1088/1748-0221/10/03/C03034}{2015 JINST 10 C03034}

\bibitem{run34note} A. Kluge, P. Vande Vyvre, \emph{The detector read-out in ALICE during Run 3 and 4}, ALICE-TECH-2016-001

\bibitem{ONU1} D.M. Kolotouros \emph{et al}, \emph{A TTC upgrade proposal using bidirectional 10G-PON FTTH technology}, \href{https://iopscience.iop.org/article/10.1088/1748-0221/10/04/C04001}{2015 JINST  10 C04001}

\bibitem{ONU2} E.B.S. Mendes \emph{et al}, \emph{The 10G TTC-PON: challenges, solutions and performance}, \mbox{\href{https://iopscience.iop.org/article/10.1088/1748-0221/12/02/C02041}{2017 JINST  12 C02041}}

\bibitem{SAMPA1} S.H.I. Barboza \emph{et al}, \emph{SAMPA chip: a new ASIC for the ALICE TPC and
MCH upgrades}, \mbox{\href{https://iopscience.iop.org/article/10.1088/1748-0221/12/02/C02041}{2016 JINST  11 C02088}}

\bibitem{SAMPA2} J. Adolfsson \emph{et al}, \emph{SAMPA Chip: the New 32 Channels ASIC for the
ALICE TPC and MCH Upgrades}, \href{https://iopscience.iop.org/article/10.1088/1748-0221/12/04/C04008}{2017 JINST  12 C04088}

\bibitem{run34paper} F. Costa \emph{et al}, \emph{The detector read-out in ALICE during Run 3 and 4}, \mbox{\href{https://iopscience.iop.org/article/10.1088/1742-6596/898/3/032011}{2017 J. Phys.: Conf. Ser. 898 032011 }}

\bibitem{rocDriverRef}  F. Costa \emph{et al}, \emph{The ReadoutCard userspace driver for the new ALICE O2 computing system}, \href{https://arxiv.org/abs/2010.16327}{arXiv:2010.16327}

\bibitem{Avalon} Avalon Interface Specifications, MNL-AVABUSREF,
 \href{https://www.intel.com/content/www/us/en/programmable/documentation/nik1412467993397.html}{Altera website}

\bibitem{gitProj} \emph{Project gitlab} \href{https://gitlab.cern.ch/alice-cru/cru-fw}{https://gitlab.cern.ch/alice-cru/cru-fw}


\end{thebibliography}
\end{document}